\documentclass{elsart}
\usepackage{epsfig}
\usepackage{natbib}

\begin{document}
\begin{frontmatter}
\title{Long-term persistence and multifractality of river runoff records: 
Detrended fluctuation studies}
\author[gi,po]{Eva Koscielny-Bunde}, \author[gi,bo,ha]{Jan W. Kantelhardt}, 
\author[mu]{Peter Braun}, \author[gi]{Armin Bunde}, and
\author[il,gi]{Shlomo Havlin} 
\address[gi]{Institut f\"ur Theoretische Physik III,
Justus-Liebig-Universit\"at, Giessen, Germany}
\address[po]{Potsdam Institute for Climate Impact Research, Potsdam, Germany}
\address[bo]{Center for Polymer Studies, Dept. of Physics,
Boston University, Boston, USA}
\address[mu]{Bayerisches Landesamt f\"ur Wasserwirtschaft, M\"unchen, Germany}
\address[il]{Minerva Center, Dept. of Physics, Bar-Ilan University, Ramat-Gan, 
Israel}
\address[ha]{present addr.:  Fachber. Physik, 
Martin-Luther-Universit\"at, Halle, Germany}
\date{January 23, 2003}
\begin{abstract}
We study temporal correlations and multifractal properties of long river 
discharge records from 41 hydrological stations around the globe.  To
detect long-term correlations and multifractal behaviour in the presence 
of trends, we apply several recently developed methods [detrended 
fluctuation analysis (DFA), wavelet analysis, and multifractal DFA] that 
can systematically detect and overcome nonstationarities in the data at 
all time scales.  We find that above some crossover time that usually is 
several weeks, the daily runoffs are long-term correlated, being 
characterized by a correlation function $C(s)$ that decays as $C(s) 
\sim s^{-\gamma}$.  The exponent $\gamma$ varies from river to river in a 
wide range between 0.1 and 0.9.  The power-law decay of $C(s)$ corresponds 
to a power-law increase of the related fluctuation function
$F_2(s)\sim s^{H}$ where $H=1-\gamma/2$.  We also find that in most 
records, for large times, weak multifractality occurs.  The Renyi exponent 
$\tau(q)$ for $q$ between $-10$ and $+10$ can be fitted to the remarkably 
simple form $\tau(q) = -\ln (a^q+b^q)/\ln 2$, with solely two parameters 
$a$ and $b$ between 0 and 1 with $a+b \ge 1$.  This type of multifractality 
is obtained from a generalization of the  multiplicative cascade model. 
\end{abstract}
\end{frontmatter}

\section{Introduction}
The analysis of river flows has a long history.  Already more than half a
century ago Hurst found by means of his $R/S$ analysis that annual runoff
records from various rivers (including the Nile river) exhibit
"long-range statistical dependencies" \citep{Hurst51}, indicating 
that the fluctuations in water storage and runoff processes are self-similar 
over a wide range of time scales, with no single characteristic scale.  
Hurst's finding is now recognized as the first example for self-affine 
fractal behaviour in empirical time series, see e.g. \citet{Feder88}.
In the 1960s, the "Hurst phenomenon" was investigated on a broader 
empirical basis for many other natural phenomena \citep{Hurst65,Mandelbrot69}.

The scaling of the fluctuations with time is reflected by the scaling of 
the power spectrum $E(f)$ with frequency $f$, $E(f) \sim f^{-\beta}$.  For 
stationary time series the exponent $\beta$ is related to the decay of the 
corresponding autocorrelation function $C(s)$ of the runoffs (see Eq. (1)).  
For $\beta$ between 0 and 1, $C(s)$ decays by a power law, $C(s) \sim 
s^{-\gamma}$, with $\gamma = 1 - \beta$ being restricted to the interval 
between 0 and 1.  In this case, the mean correlation time diverges, and the 
system is regarded as long-term correlated.  For $\beta = 0$, the runoff data 
are uncorrelated on large time scales ("white noise").  The exponents $\beta$ 
and $\gamma$ can also be determined from a fluctuation analysis, where the 
departures from the mean daily runoffs are considered as increments of a 
random walk process.  If the runoffs are uncorrelated, the fluctuation function
$F_2(s)$, which is equivalent to the root-mean-square displacement of the 
random walk, increases as the square root of the time scale $s$, $F_2(s) \sim 
\sqrt{s}$.  For long-term correlated data, the random walk becomes anomalous, 
and $F_2(s) \sim s^H$.  The fluctuation exponent $H$ is related to the 
exponents $\beta$ and $\gamma$ via $\beta = 1-\gamma = 2 H -1$.  For 
monofractal data, $H$ is identical to the classical Hurst exponent.
Recently, many studies using these kinds of methods have dealt with scaling
properties of hydrological records and the underlying statistics, see 
e.~g. \citet{Lovejoy91,Turcotte93,Gupta94,Tessier96,Davis96,Rodriguez97,
Pandey98,Matsoukas00,Montanari00,Peters02,Livina03a,Livina03b}.

However, the conventional methods discussed above may fail when trends are 
present in the system.  Trends are systematic deviations from the average runoff 
that are caused by external processes, e.g. the construction of a water 
regulation device, the seasonal cycle, or a changing climate (e.g. {\it global 
warming}).  Monotonous trends may lead to an overestimation of the Hurst 
exponent and thus to an underestimation of $\gamma$.  It is even possible 
that uncorrelated data, under the influence of a trend, look like long-term 
correlated ones when using the above analysis methods.  In addition, 
long-term correlated data cannot simply be detrended by the common 
technique of moving averages,  since this method destroys the correlations 
on long time scales (above the window size used).  Furthermore, it is 
difficult to distinguish trends from long-term correlations, because 
stationary long-term correlated time series exhibit persistent behaviour 
and a tendency to stay close to the momentary value.  This causes positive 
or negative deviations from the average value for long periods of time 
that might look like a trend. 

In the last years, several methods such as wavelet techniques (WT) and 
detrended fluctuation analysis (DFA), have been developed that are able to 
determine long-term correlations in the presence of trends.  For details 
and applications of the methods to a large number of meteorological, 
climatological and biological records we refer to 
\citet{Peng94,Taqqu95,Bunde00,Kantelhardt01,Arneodo02,Bunde02}.  The 
methods, described in Section 2, consider fluctuations in the cumulated 
runoffs (often called the "profile" or "landscape" of the record).  They 
differ in the way the fluctuations are determined  and in the type of 
polynomial trend that is eliminated in each time window of size $s$.  

In this paper, we apply these detrending methods to study the scaling 
of the fluctuations $F_2(s)$ of river flows with time $s$.  We focus 
on 23 runoff records from international river stations spread around the 
globe and compare the results with those of 18 river stations from southern 
Germany.  We find that above some crossover time (typically several weeks)
$F_2(s)$ scales as $s^{H}$ with $H$ varying from river to river between 
0.55 and 0.95 in a nonuniversal manner independent of the size of the basin. 
The lowest exponent $H=0.55$ was obtained for rivers on permafrost ground.  
Our finding is not consistent with the hypothesis that the scaling is 
universal with an exponent close to 0.75 \citep{Hurst65,Feder88} with the 
same power law being applicable for all time scales from minutes till 
centuries.  

The above detrending approaches, however, are not sufficient to fully 
characterize the complex dynamics of river flows, since they exclusively 
focus on the variance which can be regarded as the second moment $F_2(s)$ of 
the full distribution of the fluctuations.  Note that  the Hurst method 
actually focuses on the first moment $F_1(s)$. To further characterize a 
hydrological record, we extend the study to include all moments $F_q(s)$.  
A detailed description of the method, which is a multifractal generalization 
of the detrended fluctuation analysis \citep{Kantelhardt02} and equivalent 
to the Wavelet Transform Modulus Maxima (WTMM) method \citep{Arneodo02}, 
is given in Section 3.  Our approach differs from the multifractal approach 
introduced into hydrology by Lovejoy and Schertzer (see e.g. 
\citet{Schertzer87,Lovejoy91,Lavallee93,Pandey98}) that was based on the 
concept of structure functions \citep{Frisch85} and on the assumption 
of the existence of a universal cascade model.  Here we perform the 
multifractal analysis by studying how the different moments of the 
fluctuations $F_q(s)$ scale with time $s$, see also \citet{Rodriguez97}.  
We find that at large time scales, $F_q(s)$ scales as $s^{h(q)}$, and a 
simple functional form with two parameters ($a$ and $b$), $h(q) = (1/q) - 
[\ln(a^q + b^q)]/[q\ln(2)]$ describes the scaling exponent $h(q)$ of all 
moments.  On small time scales, however, a stronger multifractality is 
observed that may be partly related to the seasonal trend.  The mean 
position of the crossover between the two regimes is of the order of weeks 
and increases with $q$.

\section{Correlation Analysis}

Consider a record of daily water runoff values $W_i$ measured 
at a certain hydrological station. The index $i$ counts the days in the
record, $i=1$,2,\ldots,$N$. To eliminate the periodic seasonal trends,
we concentrate on the departures $\phi_i=W_i - \overline{W_i}$ from the 
mean daily runoff $\overline{W_i}$. $\overline{W_i}$ is calculated for each
calendar date $i$ (e.g. April 1$^{st}$) by averaging over all years in the 
runoff series.  In addition, we checked that our actual results remained 
unchanged when also seasonal trends in the variance have been eliminated by 
analysing $\phi'_i=(W_i - \overline{W_i}) / (\overline{W^2_i} - 
\overline{W_i}^2 )^{1/2}$ instead of $\phi_i$. 

The runoff autocorrelation function $C(s)$ describes, how the persistence 
decays in time.  If the $\phi_i$ are uncorrelated, $C(s)$ is zero for all 
$s$.  If correlations exist only up to a certain number of days $s_\times$, 
the correlation function will vanish above $s_\times$.  For long-term 
correlations, $C(s)$ decays by a power-law 
\begin{equation} C(s)=\langle  \phi_i\phi_{i+s}\rangle 
\sim s^{-\gamma}, \qquad 0<\gamma<1, \label{gamma}
\end{equation}
where the average $\langle \ldots \rangle$ is over all pairs with the same 
time lag $s$.  For large values of $s$, a direct calculation of $C(s)$ is 
hindered by the level of noise present in the finite hydrological records, 
and by nonstationarities in the data.  There are several alternative methods 
for calculating the correlation function in the presence of long-term 
correlations, which we describe in the following sections.

\subsection{Power Spectrum Analysis}

If the time series is  stationary, we can apply standard spectral
analysis techniques and  calculate the power spectrum $E(f)$ of the time
series $W_i$ as a function of the frequency $f$.  For long-term correlated 
data, we have $E(f) \sim f^{-\beta},$ where $\beta$ is related to the 
correlation exponent $\gamma$ by $\beta= 1-\gamma$.  This relation can be 
derived from the Wiener-Khinchin theorem.  If, instead of $W_i$ the 
integrated runoff time series $z_n = \sum_{i=1}^n \phi_i$ is Fourier 
transformed, the resulting power spectrum scales as $\tilde{E}(f) \sim 
f^{-2-\beta}$.

\subsection{Standard Fluctuation Analysis (FA)}

In the standard fluctuation analysis,  we consider the ``runoff profile"
\begin{equation} z_n=\sum_{i=1}^n \phi_i, \qquad n=1,2,\ldots,N,
\label{profile}\end{equation}
and study how the fluctuations of the profile, in a given time window 
of size $s$, increase with $s$.  We can consider the profile $z_n$ as 
the position of a random walker on a linear chain after $n$ steps. 
The random walker starts at the origin and performs, in the $i$th step, 
a jump of length $\phi_i$ to the right, if  $\phi_i$ is positive, and 
to the left, if  $\phi_i$ is negative.

To find how the square-fluctuations of the profile scale with $s$, we
first divide each record of $N$ elements into $N_s= {\rm int}(N/s)$ 
nonoverlapping segments of size $s$ starting from the beginning and 
$N_s$ nonoverlapping segments of size $s$ starting from the end of the 
considered runoff series. Then we determine the fluctuations in each 
segment $\nu$.

In the standard fluctuation analysis, we obtain  the fluctuations just 
from the values of the profile at both endpoints of each segment $\nu$,
$F^2(\nu,s)=[z_{\nu s}-z_{(\nu-1) s}]^2$, and average $F^2(\nu,s)$ over 
the $2N_s$ subsequences to obtain the mean fluctuation $F_2(s)$,
\begin{equation} F_2(s) \equiv \left\{ {1 \over 2 N_s} \sum_{\nu=1}^{2 N_s} 
F^2(\nu,s)\right\}^{1/2}. \label{Fs2}\end{equation}
By definition, $F_2(s)$ can be viewed as the 
root-mean-square displacement of the random walker on the chain, after 
$s$ steps.  For uncorrelated $\phi_i$ values, we obtain Fick's diffusion 
law $F_2(s) \sim s^{1/2}$.  For the relevant case of long-term correlations,  
where $C(s)$ follows the power-law behaviour of Eq. (1), $F_2(s)$ increases 
by a power law (see, e.g., \citet{Bunde02}),
\begin{equation} F_2(s) \sim s^{H}, \label{H}\end{equation}
where the fluctuation exponent $H$ is related to the correlation exponent 
$\gamma$ and the power-spectrum exponent $\beta$ by
\begin{equation}  H=1-\gamma/2=(1+\beta)/2. \label{HH}\end{equation}
For power-law correlations decaying faster than $1/s$, we have $H=1/2$ 
for large $s$ values, like for uncorrelated data. 

We like to note that the standard fluctuation analysis is somewhat similar 
to the rescaled range analysis introduced by Hurst (for a review see, e.~g., 
\citet{Feder88}), except that it focusses on the second moment $F_2(s)$ 
while Hurst considered the first moment $F_1(s)$. For monofractal data, 
$H$ is identical to the Hurst exponent.

\subsection{The Detrended Fluctuation Analysis (DFA)}

There are different orders of DFA that are distinguished by the way the
trends in the data are eliminated. In lowest order (DFA1) we determine,
for each segment $\nu$, the best {\it linear} fit  of the profile, and
identify the fluctuations by the variance $F^2(\nu,s) $ of the profile
from this straight line.  This way, we eliminate the influence of
possible linear trends on scales larger than the segment. Note that
linear trends in the profile correspond to patch-like trends in the
original record.  DFA1 has been proposed originally by \citet{Peng94} 
when analyzing correlations in DNA. It can be generalized
straightforwardly to eliminate higher order trends 
\citep{Bunde00,Kantelhardt01}.

In second order DFA (DFA2) one calculates the variances $F^2(\nu,s) $ of
the profile from best {\it quadratic} fits of the profile, this way
eliminating the influence of possible linear and parabolic trends on
scales larger than the segment considered.  In general, in $n$th-order DFA, 
we calculate the variances of the profile from the best $n$th-order 
polynomial fit, this way eliminating the influence of possible 
$(n-1)$th-order trends on scales larger than the segment size.

Explicitly, we calculate the best polynomial fit $y_\nu (i)$ of the 
profile in each of the $2 N_s$ segments $\nu$ and determine the variance
\begin{equation} F^2(\nu,s) \equiv {1 \over s} \sum_{i=1}^{s} 
\left[ z_{(\nu-1) s + i} - y_{\nu}(i) \right]^2. \label{DFA1} 
\end{equation}
Then we employ Eq.~(\ref{Fs2}) to determine the mean fluctuation 
$F_2(s)$.

Since FA and the various stages of the DFA have different detrending 
capabilities, a comparison of the fluctuation functions obtained by 
FA and DFA$n$ can yield insight into both long-term correlations and
types of trends.  This cannot be achieved by the conventional methods, 
like the spectral analysis.

\subsection{Wavelet Transform (WT)}

The wavelet methods we employ here are based on the determination of 
the mean values $\overline z_{\nu}(s)$ of the profile in each segment 
$\nu$ (of length $s$), and the calculation of the fluctuations between 
neighbouring segments.  The  different order techniques we have used in 
analyzing runoff fluctuations differ in the way the fluctuations 
between the average profiles are treated and possible nonstationarities 
are eliminated. The first-, second- and third-order wavelet method are 
described below.

(i) In the first-order wavelet method (WT1), one simply determines the
fluctuations from the first derivative $F^2(\nu,s)=[\overline z_\nu(s)-
\overline z_{\nu+1}(s)]^2$.  WT1 corresponds to FA where constant trends 
in the profile of a hydrological station are eliminated, while linear 
trends are not eliminated.

(ii) In the second-order wavelet method (WT2), one determines the
fluctuations from the second derivative $F^2(\nu,s)=[\overline z_\nu(s)-
2\overline z_{\nu+1}(s)+ \overline z_{\nu+2}(s)]^2$.  So, if the profile 
consists of a trend term linear in $s$ and a fluctuating term, the trend 
term is eliminated. Regarding trend-elimination, WT2 corresponds to DFA1.

(iii) In the third-order wavelet method (WT3), one determines the
fluctuations from the third derivative $F^2(\nu,s)=[\overline z_\nu(s)-
3\overline z_{\nu+1}(s) + 3\overline z_{\nu+2}(s)-\overline z_{\nu+3}(s)]^2$.
By definition, WT3 eliminates linear and parabolic trend terms in the
profile.  In general, in WT$n$ we determine the fluctuations from the
$n$th derivative, this way eliminating trends described by 
$(n-1)$st-order polynomials in the data. 

Methods (i-iii) are called wavelet methods, since they can be interpreted
as transforming the profile by discrete wavelets representing first-,
second- and third-order cumulative derivatives of the profile. The
first-order wavelets are known in the literature as Haar wavelets. One can
also use different shapes of the wavelets (e.g. Gaussian wavelets with 
width $s$), which have been used by \citet{Arneodo02} to study, 
for example, long-range correlations in DNA.  Since the various 
stages of the wavelet methods WT1, WT2, WT3, etc. have different detrending 
capabilities, a comparison of their fluctuation functions can yield insight 
into both long-term correlations and types of trends.

At the end of this section, before describing the results of the FA, DFA, 
and WT analysis, we note that for very large $s$ values, $s>N/4$ for DFA
and $s>N/10$ for  FA and WT, the fluctuation function becomes inaccurate 
due to statistical errors. The difference in the statistics is due to the 
fact that the number of independent segments of length $s$ is larger in 
DFA than in WT, and the fluctuations in FA are larger than in DFA.   Hence, 
in the analysis we will concentrate on $s$ values lower than $s_{\rm max} = 
N/4$ for DFA and $s_{\rm max} = N/10$ for FA and WT.  When determining the 
scaling exponents $H$ using Eq.~(\ref{H}), we manually chose an appropriate 
(shorter) fitting range of typically  two orders of magnitude.

\subsection{Results} \label{results}

\begin{figure} 
\noindent\centering\epsfig{file=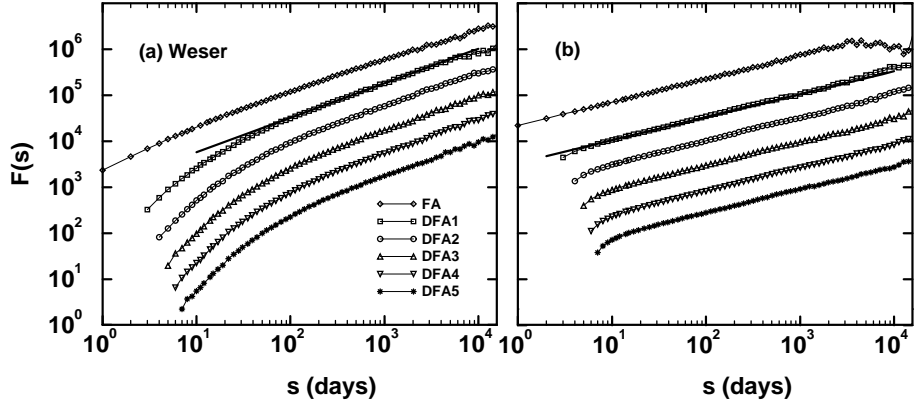,width=12cm}
\caption{(a) The fluctuation functions $F_2(s)$ versus time scale $s$ 
obtained from FA and DFA1-DFA5 in double logarithmic plots for daily 
runoff departures $\phi_i=W_i - \overline{W_i}$ from the mean daily 
runoff $\overline{W_i}$ for the river Weser, measured from 1823 till 
1993 by the hydrological station Vlotho in Germany.  (b) The analog 
curves to (a) when the $\phi_i$ are randomly shuffled.} \end{figure}

In our study we analyzed 41 runoff records, 18 of them are from the 
southern part of Germany, and the rest is from North and South America, 
Africa, Australia, Asia and Europe (see Table 1). We begin the analysis 
with the runoff record for the river Weser in the northern part of 
Germany, which has the longest record (171 years) in this study.  Figure~1(a) 
shows the fluctuation functions $F_2(s)$ obtained from FA and DFA1-DFA5.  
In the log-log plot, the curves are approximately straight lines for $s$ 
above 30 days, with a slope $H \approx 0.75$.  This result for the 
Weser suggests that there exists long-term persistence expressed by the 
power-law decay of the correlation function, with an exponent $\gamma 
\approx 0.5$ [see Eq.~(\ref{H})]. 

To show that the slope $H \approx 0.75$ is due to long-term correlations 
and not due to a broad probability distribution (Joseph- versus 
Noah-Phenomenon, see \citet{Mandelbrot68}), we have eliminated the 
correlations by randomly shuffling the $\phi_i$.  This shuffling has no 
effect on the probability distribution function of $\phi_i$.   Figure 
1(b) shows $F_2(s)$ for the shuffled data.  We obtain $H=1/2$, showing 
that the exponent $H \approx 0.75$ is due to long-term correlations.

\begin{figure}
\noindent\centering\epsfig{file=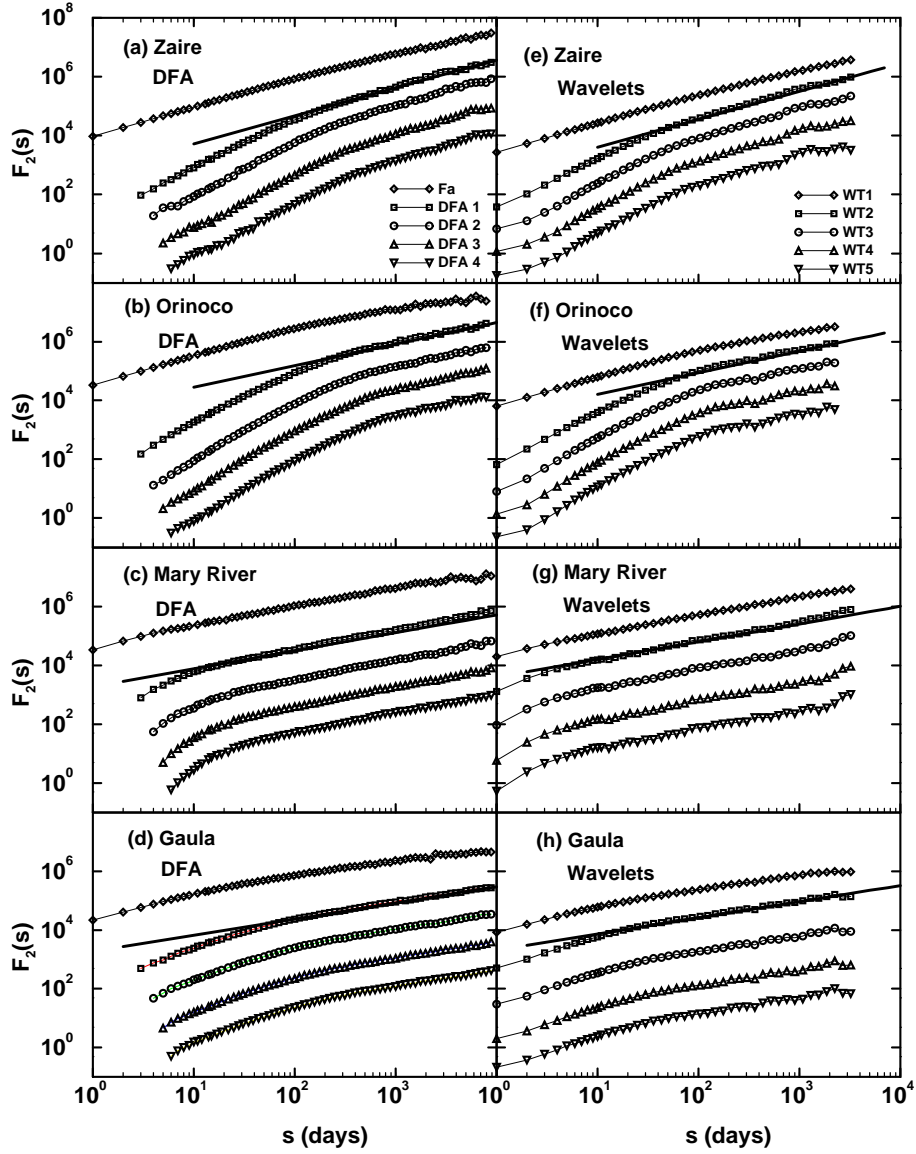,width=12cm}
\caption{The fluctuation functions $F_2(s)$ versus time scale $s$ 
obtained from FA and DFA1-DFA4 in double logarithmic plots for four 
additional representative hydrological stations: (a) the Zaire in 
Kinshasa, Zaire, (b) the Orinoco in Puente Angostura, Venezuela,
(c) the Mary River in Miva, Australia, and (d) the Gaula River in Haga Bru, 
Norway.  (e-h) the fluctuation functions $F_2(s)$ obtained for the same 
rivers as in (a-d), from first to fifth order wavelet analysis 
(WT1-WT5).  The straight lines are best linear fits to the DFA1 and the WT2 
results on large time scales.} \end{figure}

To show that the slope $H \approx 0.75$ is not an artefact of the 
seasonal dependence of the variance and skew, we also considered records 
where $\phi_i$ was divided by the variance of each calendary day and 
applied further detrending techniques that take into account the skew 
\citep{Livina03b}.  In both cases, we found no change in the 
scaling behaviour for large times (see also Sect. 3.4).  This can be 
understood easily, since kinds of seasonal trends cannot effect the 
fluctuation behaviour on time scales well above one year.  It is likely, 
however, that the seasonal dependencies of the variance and 
possibly also of the skew contribute to the behaviour at small times,
where the slope $H$ is much larger than 0.75 in most cases (see also
Sect. 3.4, where a seasonal trend in the variance is indeed used for 
modelling the crossover).

Figure 2 shows the fluctuation functions $F_2(s)$ of 4 more rivers, from 
Africa, South America, Australia, and Europe.  The panels on the left-hand 
side show the FA and DFA1-4 curves, while the panels on the right-hand 
side show the results from the analogous wavelet analysis WT1-WT5.  Most 
curves show a crossover at small time scales; a similar crossover has been 
reported by \citet{Tessier96} for small French rivers without artificial 
dams or reservoirs.  Above the crossover time, the fluctuation functions 
(from DFA1-4 and WT2-5) show power-law behaviour, with exponents $H \simeq 
0.95$ for the Zaire, $H \simeq 0.73$ for the Orinoco, $H \simeq 0.60$ for 
the Mary river, and $H \simeq 0.55$ for the Gaula river.  

Accordingly, there is no universal scaling behaviour since the long-term 
exponents vary strongly from river to river and reflect the fact that 
there exist different mechanisms for floods where each may induce different 
scaling.  This is in contrast to climate data, where universal long-term 
persistence of temperature records at land stations was observed 
\citep{Koscielny98,Talkner00,Weber01,Eichner03}. 

The Mary river in Australia is rather dry in the summer and the Gaula 
river in Norway is frozen in the winter.  For the Mary river, the 
long-term exponent $H \simeq 0.60$ is well below the average value.  
For the Gaula river, the long-term correlations are not pronounced 
($H=0.55$) and even hard to distinguish from the uncorrelated case 
$H=0.5$.  We obtained similar results for the other two "frozen" 
rivers (Tana from Norway and Dvina from Russia) that we analysed.  
For interpreting this distinguished behaviour of the frozen rivers 
we like to note that on permafrost ground the lateral inflow (and hence 
the indirect contribution of the water storage in the catchment basin) 
contributes to the runoffs in a different way than on normal ground, see 
also {\it Gupta and Dawdy} [1995].  Our results (based on 3 rivers only) 
suggest that the contribution of snow melting leads to less correlated 
runoffs than the contribution of rainfall, but  more comprehensive 
studies will be needed to confirm this interesting result.

Figure 3(a) and Table 1 summarize our results for $H$.  One can see 
clearly that the exponents $H$ do not depend systematically on the basin 
area $A$.  This is in line with the conclusions of \citet{Gupta94} for 
the flood peaks, where a systematic dependence on $A$ could also not be 
found.

\begin{table}
\noindent\centering\epsfig{file=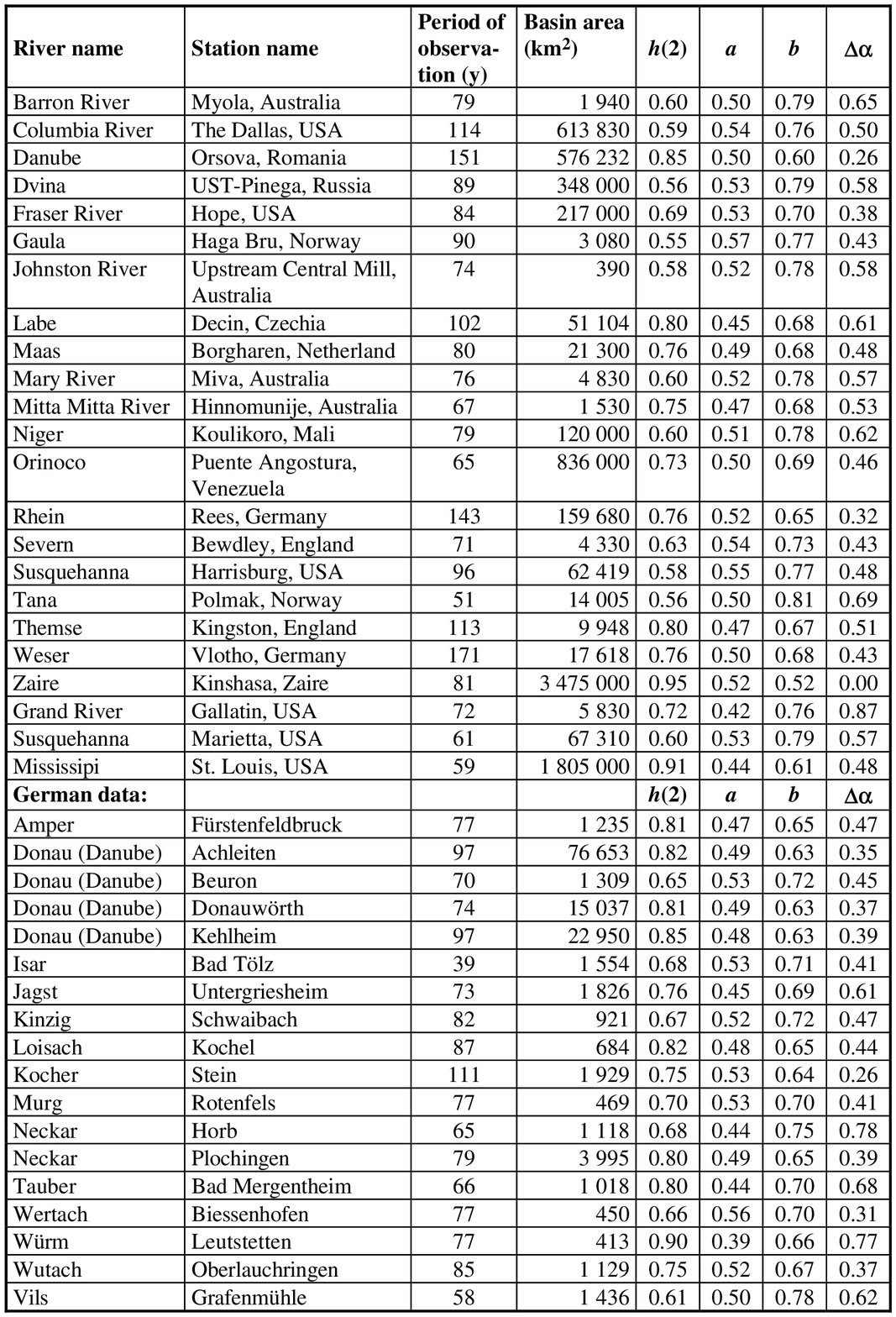,width=13cm,height=19cm}
\caption{Table of investigated international river basins (data 
from Global Runoff Data Center (GRDC), Koblenz, Germany) and investigated 
South German river basins.  We list the river and station name, the 
duration of the investigated daily record, the size of the basin area, 
and the results of our analysis, $H=h(2)$, the multifractal quantities 
$a$, $b$, and $\Delta \alpha$.} \end{table}

\begin{figure}
\noindent\centering\epsfig{file=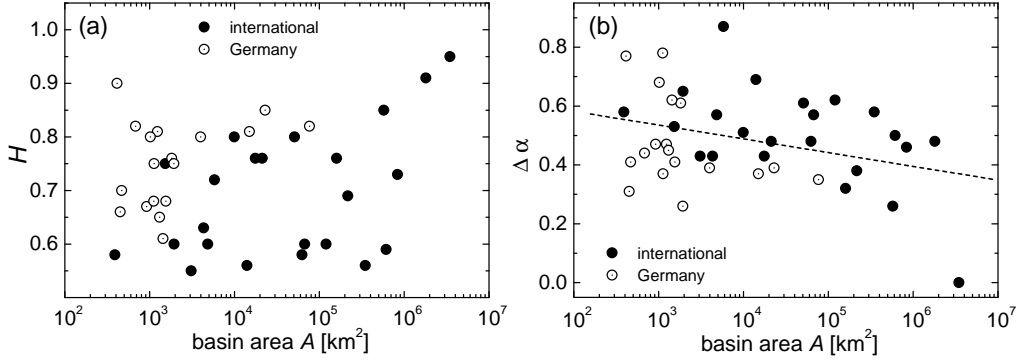,width=4.7cm,angle=-90}
\caption{(a) Long-term fluctuation exponents $H$ and (b) widths 
$\Delta\alpha$ of the $f(\alpha)$ spectra for all international records 
(full symbols) and all records from south Germany (open symbols) that we 
analyzed, as a function of the basin area $A$.  Each symbol represents the 
result for one hydrological station.  The dashed line in (b) is a linear 
fit to the data.} \end{figure}

There is also no pronounced regional dependence: the rivers within a 
localized area (such as South Germany) tend to have nearly the same range 
of exponents as the international rivers.  The three ''frozen'' rivers in 
our study have the lowest values of $H$.  As can be seen in the figure, the 
exponents spread from 0.55 to 0.95.  Since the correlation exponent $\gamma$ 
is related to $H$ by $\gamma = 2 - 2H$, the exponent $\gamma$ spreads from 
almost 0 to almost 1, covering the whole range from very weak to very strong 
correlations.

\section{Multifractal Analysis}

\subsection{Method}

For a further characterization of hydrological records it is meaningful 
to extend Eq.~(\ref{Fs2}) by considering the more general fluctuation 
functions (\citet{Barabasi91}, see also \citet{Davis96}), 
\begin{equation} F_q(s) = \left\{ {1 \over 2 N_s} \sum_{\nu=1}^{2 N_s} 
\vert z_{\nu s} - z_{(\nu-1) s} \vert^q \right\}^{1/q}. 
\label{FAq} \end{equation}
where the variable $q$ can take any real value except zero.  For $q=2$, 
the standard fluctuation analysis is 
retrieved.  The question is, how the fluctuation functions depend on $q$ 
and how this dependence is related to multifractal features of the record.

In general, the multifractal approach is introduced by the partition 
function
\begin{equation} Z_q(s) \equiv \sum_{\nu=1}^{N_s} \vert z_{\nu s} -
z_{(\nu-1) s} \vert^q \sim s^{\tau(q)}, \label{Zq} \end{equation}
where $\tau(q)$ is the Renyi scaling exponent.  A record is called 
'monofractal', when $\tau(q)$ depends linearly on $q$; otherwise it 
is called multifractal.

It is easy to verify that $Z_q(s)$ is related to $F_q(s)$ by
\begin{equation} F_q(s)=\left\{ {1 \over N_s} Z_q(s) \right\}^{1/q}. 
\label{FZ} \end{equation}
Accordingly, Eq.~(\ref{Zq}) implies
\begin{equation} F_q(s)\sim s^{h(q)}, \label{Fqs} \end{equation}
where
\begin{equation} h(q)= [\tau(q) + 1]/q. \label{htau} \end{equation}
Thus, $h(q)$ defined in Eq.~(\ref{Fqs}) is directly related to the 
classical multifractal scaling exponents $\tau(q)$. 

In general, the exponent $h(q)$ may depend on $q$.  Since for 
stationary records, $h(1)$ is identical to the well-known Hurst 
exponent (see e.~g. \citet{Feder88}), we will call the 
function $h(q)$ the generalized Hurst exponent.  For monofractal 
self-affine time series, $h(q)$ is independent of $q$, since the
scaling behaviour of the variances $F^2(\nu,s)$ is identical for all
segments $\nu$, and the averaging procedure in Eq.~(\ref{FAq}) will
give just this identical scaling behaviour for all values of $q$. 
If small and large fluctuations scale differently, there will be a
significant dependence of $h(q)$ on $q$:  If we consider positive
values of $q$, the segments $\nu$ with large variance $F^2(\nu,s)$
(i.~e. large deviations from the corresponding fit) will dominate 
the average $F_q(s)$.  Thus, for positive values of $q$, $h(q)$
describes the scaling behaviour of the segments with large
fluctuations.  Usually the large fluctuations are characterized 
by a smaller scaling exponent $h(q)$ for multifractal series. 
On the contrary, for negative values of $q$, the segments $\nu$ with
small variance $F^2(\nu,s)$ will dominate the average $F_q(s)$. 
Hence, for negative values of $q$, $h(q)$ describes the scaling
behaviour of the segments with small fluctuations, which are usually
characterized by a larger scaling exponent.

In the hydrological literature \citep{Rodriguez97,Lavallee93} one 
often considers the generalized mass variogram $C_q(\lambda)$
(see also \citet{Davis96}),
\begin{equation} 
C_q(\lambda) \equiv \langle \vert z_{i+\lambda} - z_i \vert^q \rangle
\sim \lambda^{K(q)}. \label{Cq}\end{equation}
Comparing Eqs.~(\ref{FAq}), (\ref{Fqs}), and (\ref{Cq}) one can verify 
easily that $K(q)$ and $h(q)$ are related by
\begin{equation} h(q) = K(q)/q. \label{ksh} \end{equation}

Another way to characterize a multifractal series is the singularity
spectrum $f(\alpha)$, that is related to $\tau(q)$ via a Legendre
transform (e.g. \citet{Feder88,Rodriguez97}),
\begin{equation} \alpha = {d \tau(q) \over dq }\quad {\rm and} \quad
f(\alpha) = q \alpha - \tau(q). \label{Legendre} \end{equation}
Here, $\alpha$ is the singularity strength or H\"older exponent, while
$f(\alpha)$ denotes the dimension of the subset of the series that is
characterized by $\alpha$. Using Eq.~(\ref{htau}), we can directly
relate $\alpha$ and $f(\alpha)$ to $h(q)$, 
\begin{equation} \alpha = h(q) + q {dh(q) \over dq} \quad {\rm and}
\quad f(\alpha) = q [\alpha - h(q)] + 1.\label{Legendre2} \end{equation}
The strength of the multifractality of a time series can be 
characterized by the difference between the maximum and minimum 
values of $\alpha$, $\alpha_{\rm max} - \alpha_{\rm min}$. When 
$q {dh(q) \over dq}$ approaches zero for $q$ approaching $\pm\infty$, 
then  $\Delta\alpha=\alpha_{\rm max} - \alpha_{\rm min}$ is simply 
given by $\Delta\alpha=h(-\infty)-h(\infty)$.

The multifractal analysis described above is a straightforward 
generalization of the fluctuation analysis and therefore has the same 
problems: (i) monotonous trends in the record may lead to spurious results 
for the fluctuation exponent $h(q)$ which in turn leads to spurious results 
for the correlation exponent $\gamma$, and  (ii) nonstationary behaviour 
characterized by exponents $h(q) > 1$ cannot be detected by the simple 
method since the method cannot distinguish between exponents $>1$, and 
always will yield $F_2(s) \sim s$ in this case (see above).  To overcome 
these drawbacks the multifractal detrended fluctuation analysis (MF-DFA) 
has been introduced recently (\citet{Kantelhardt02}, see also 
\citet{Koscielny98,Weber01}).  According to \citet{Kantelhardt02,Kantelhardt03}, 
the method is as accurate as the wavelet methods.  Thus, we have used MF-DFA 
for the multifractal analysis here.  In the MF-DFA, one starts with the 
DFA-fluctuations $F_\nu^2(s)$ as obtained in Eq. (\ref{DFA1}).  Then, we 
define in close analogy to Eqs. (\ref{Fs2}) and (\ref{FAq}) the generalized 
fluctuation function,
\begin{equation} F_q(s) \equiv \left\{ {1 \over 2 N_s} \sum_{\nu=1}^{2 N_s} 
\left[F^2(\nu,s)\right]^{q/2}\right\}^{1/q}. \label{Fsq}\end{equation}
Again, we can distinguish MF-DFA1, MF-DFA2, etc., according to the order 
of the polynomial fits involved.

\begin{figure}
\noindent\centering\epsfig{file=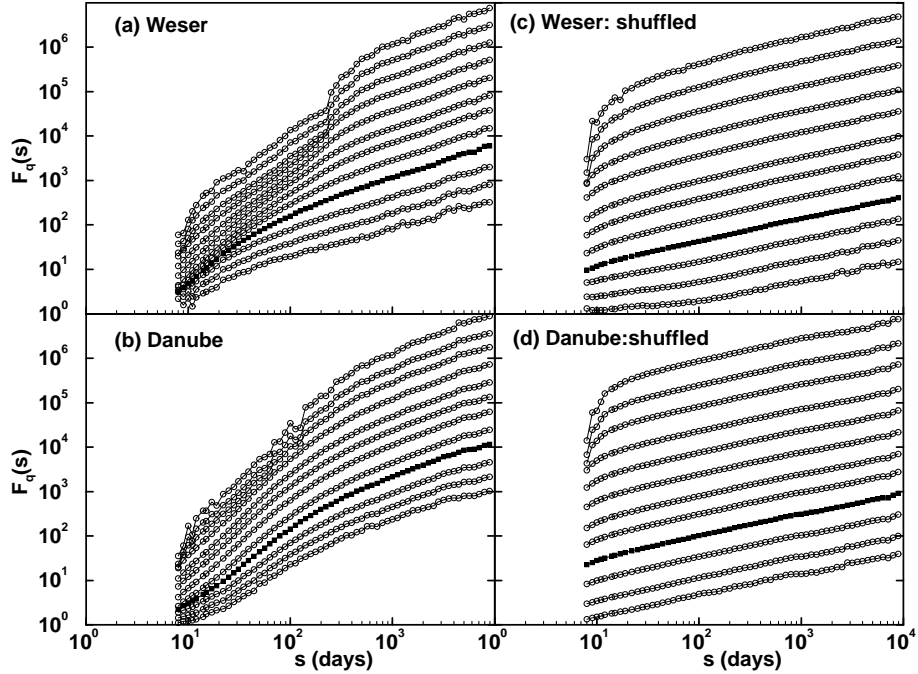,height=12cm,angle=-90}
\caption{The multifractal fluctuation functions $F_q(s)$ versus time 
scale $s$ obtained from multifractal DFA4 for two representative 
hydrological stations: (a) river Weser in Vlotho, Germany and (b)
river Danube in Orsova, Romania.  The curves correspond to different
values of $q$, $q=-10$, -6, -4, -2, -1, -0.2, 0.2, 1, 2, 4, 6, 10 (from 
the top to the bottom) and are shifted vertically for clarity. }
\end{figure}

\subsection{Multifractal Scaling Plots}

We have performed a large scale multifractal analysis on all 41 rivers. We 
found that MF-DFA2-5 yield similar results for the fluctuation function 
$F_q(s)$. We have also cross checked the results  using the Wavelet Transform 
Modulus Maxima (WTMM) method \citep{Arneodo02}, and always find agreement 
within the error bars \citep{Kantelhardt03}. 
Therefore, we present here only the results of MF-DFA4.  
Figure 4(a,b) shows two representative examples for the fluctuation functions 
$F_q(s)$, for (a) the Weser river and (b) the Danube river.  The standard 
fluctuation function $F_2(s)$ is plotted in full symbols.  The crossover 
in $F_2(s)$ that was discussed in Sect.~\ref{results} can be also seen in 
the other moments.  The position of the crossover increases monotonously 
with decreasing $q$ and the crossover becomes more pronounced.   We are only
interested in the asymptotic behaviour of $F_q(s)$ at large times $s$. One 
can see clearly that above the crossover, the $F_q(s)$ functions are straight 
lines in the double logarithmic plot, and the slopes increase slightly when 
going from high positive moments towards high negative moments (from the 
bottom to the top).  For the Weser, for example, the slope changes from 0.65 
for $q=10$ to 0.9 for $q=-10$ (see also Fig. 6(b)). The monotonous increase of 
the slopes, $h(q)$, is the signature of multifractality.

\begin{figure}
\noindent\centering\epsfig{file=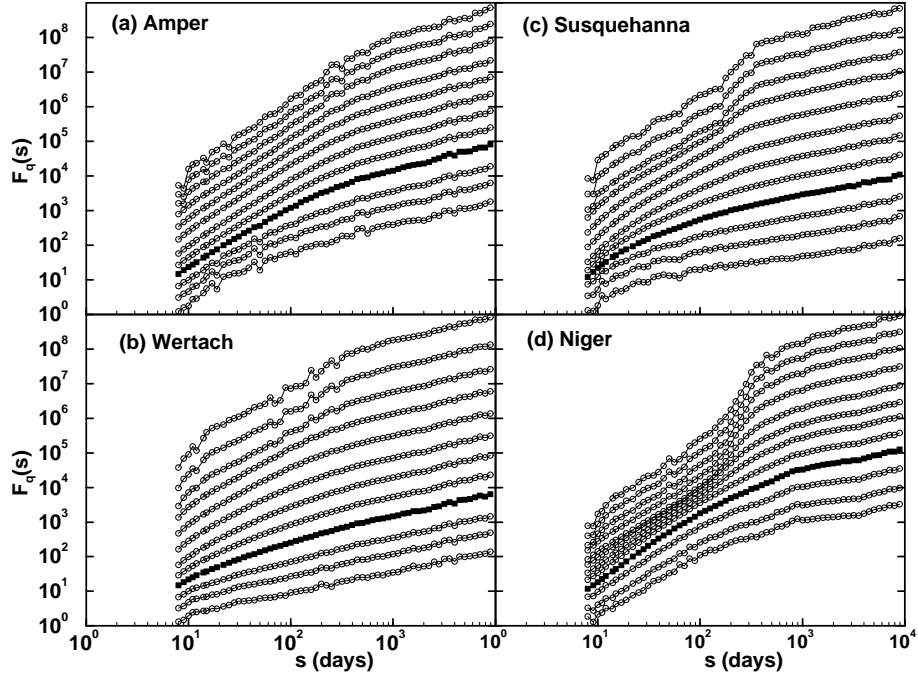,height=12cm,angle=-90}
\caption{The multifractal fluctuation functions $F_q(s)$ obtained from 
multifractal DFA4 for four additional hydrological stations: (a) Amper
in F\"urstenfeldbruck, Germany, (b) Wertach in Biessenhofen, Germany, (c) 
Susquehanna in Harrisburg, USA, (d) Niger in Koulikoro, Mali.  The $q$-values 
are identical to those used in Fig. 4.}	\end{figure}

When the data are shuffled (see Figs. 4(c,d)), all functions $F_q(s)$ 
increase asymptomatically as $F_q(s)\,\sim\,s^{1/2}$.  This indicates 
that the multifractality vanishes under shuffling.  Accordingly the 
observed multifractality originates in the long-term correlations of the 
record and is not caused by singularities in the distribution of the 
daily runoffs (see also \citet{Mandelbrot68}).  A reshuffling-resistent
multifractality would indicate a 'statistical' type of non-linearity 
\citep{Sivapalan02}.  We obtain similar patterns for all rivers.  Figure 5 
shows four more examples; Figs. 5(a,b) are for two rivers (Amper and Wertach) 
from southern Germany, while Figs. 5(c,d) are for Niger and Susquehama 
(Koulikoro, Mali and Harrisburg, USA). 

From the asymptotic slopes of the curves in Figs. 4(a,b) and 5(a-d), we 
obtain the generalized Hurst-exponents $h(q)$, which are plotted in Fig. 6 
(circles).  One can see that in the whole $q$-range the exponents can be 
fitted well by the formula
\begin{equation} h(q) = {1 \over q} - {\ln[a^q + b^q] \over q\ln 2},
\label{Hbin2} \end{equation}
or
\begin{equation} K(q) = 1+\tau(q) = 1 - {\ln[a^q + b^q] \over \ln 2}. 
\label{Kbin} \end{equation}
The formula can be derived from a modification of the multiplicative cascade 
model that we describe in Sect.~\ref{model}. Here we use the formula only 
to show that the infinite number of exponents $h(q)$ can be described by 
only two independent parameters, $a$ and $b$. These two parameters can then 
be regarded as multifractal finger print for a considered river. This is 
particularly important when checking models for river flows. Again, we 
like to emphasize, that these parameters have been obtained from the 
asymptotic part of the generalized fluctuation function, and are therefore 
not affected by seasonal dependencies.

\begin{figure}
\noindent\centering\epsfig{file=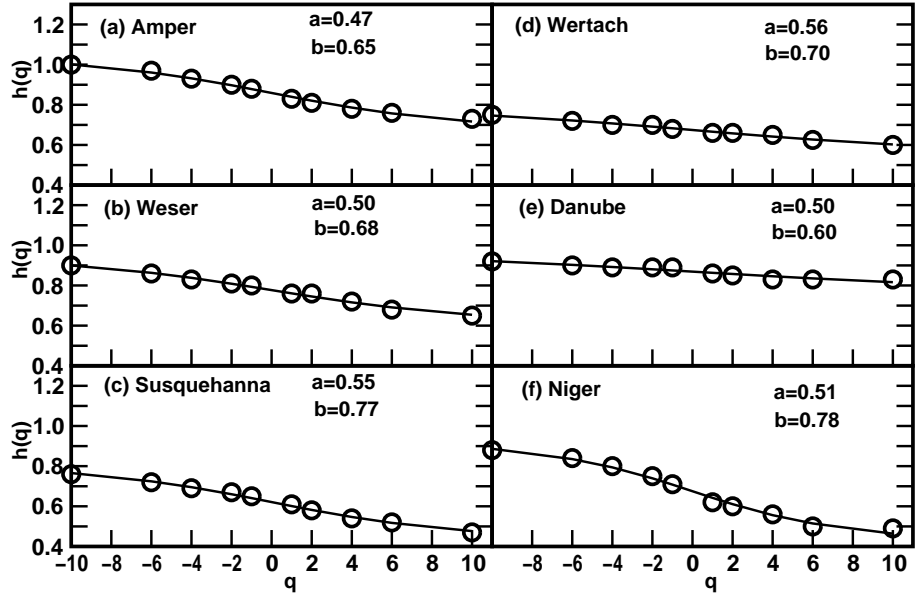,height=12cm,angle=-90}
\caption{The generalized Hurst exponents $h(q)$ for the six representative 
daily runoff records analyzed in Figs. 4 and 5: (a) Amper in 
F\"urstenfeldbruck, Germany, (b) Weser in Vlotho, Germany, (c) Susquehanna 
in Harrisburg, USA, (d) Wertach in Biessenhofen, Germany, (e) Danube in Orsova, 
Romania, and (f) Niger in Koulikoro, Mali.  The $h(q)$ values have been 
determined by straight line fits of $F_q(s)$ on large time scales.  The error 
bars of the fits correspond to the size of the symbols.} \end{figure}

We have fitted the $h(q)$ spectra in the range $-10 \le q \le 10$ for all 
41 runoff series by Eq.~(\ref{Hbin2}).  Representative examples are shown 
in Fig. 6.  The continuous lines in Fig. 6 are obtained by best fits of $h(q)$ 
(obtained from Figs. 4 and 5 as described above) by Eq.~(\ref{Hbin2}). The 
respective parameters $a$ and $b$ are listed inside the panels of each figure. 
Our results for the 41 rivers  are shown in Table 1.  It is remarkable that 
in each single case, the $q$ dependence of $h(q)$ for positive and negative 
values of $q$ can be characterized well by the two parameters, and all fits 
remain within the error bars of the $h(q)$ values. 

From $h(q)$ we obtain $\tau(q)$ (Eq.~(\ref{htau})) and the singularity 
spectrum $f(\alpha)$ (Eq.~(\ref{Legendre2})).  Figure 7 shows two typical 
examples for the Danube and the Niger.  The width of $f(\alpha)$ taken at 
$f=0$ characterizes the strength of the multifractality.  Since both widths
are very different, the strength of the multifractality of river runoffs 
appears to be not universal.

\begin{figure}
\noindent\centering\epsfig{file=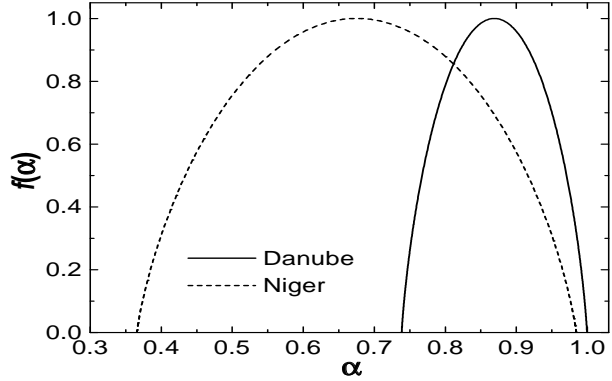,width=8.0cm,height=5.0cm}
\caption{The multifractal spectra $f(\alpha)$ for two representative 
runoff records (a) the Danube in Orsova, Romania and (b) Niger in Koulikoro, 
Mali.} \end{figure}

In order to characterize and to compare the strength of the multifractality 
for several time series we use as a parameter the width of the singularity 
spectrum $f(\alpha)$ [see Eqs.~(\ref{Legendre}) and (\ref{Legendre2})] at 
$f=0$, which corresponds to the difference of the maximum and the minimum 
value of $\alpha$.  In the multiplicative cascade model, this parameter 
is given by
\begin{equation} \Delta\alpha = {\ln a - \ln b \over \ln 2}.
\label{bindalpha} \end{equation}
The distribution of the $\Delta \alpha$ values we obtained from 
Eq.~(\ref{bindalpha}) is shown in Fig. 3(b), where we plot $\Delta \alpha$ 
versus the basin area.  The figure shows that there are rivers with quite 
strong multifractal fluctuations, i.e. large $\Delta\alpha$, and one with 
almost vanishing multifractality, i.e. $\Delta\alpha \approx 0$. Two 
observation can be made from the figure: (1) There is no pronounced 
difference between the width of the distribution of the multifractality 
strength for the runoffs within the local area of southern Germany
(open symbols in Fig. 3(b)) and for the international runoff records from 
all rivers around the globe (full symbols).  In fact, without rivers 
Zaire and Grand River the widths would be the same. (2) There is a 
tendency towards smaller multifractality strengths at larger basin 
areas. This means that the river flows become less nonlinear with 
increasing basin area. We consider this as possible indication of river 
regulations that are more pronounced for large river basins.

Our results for $K(q) = 1+\tau(q)$ (see Eq.~(\ref{Kbin}) and Table 1) may 
be compared with the functional form 
\begin{equation} K(q) = (H'+1) q - {C_1 \over \alpha'-1} (q^{\alpha'} - q)
\qquad q \ge 0 \label{lovesch} \end{equation}
with the three parameters $H'$, $C_1$, and $\alpha'$, that have been used 
by Lovejoy, Schertzer, and coworkers 
\citep{Schertzer87,Lovejoy91,Lavallee93,Tessier96,Pandey98} successfully 
to describe the multifractal behaviour of rainfall and runoff records.  
The definition of $K(q)$ we used in this paper is taken from 
\citet{Rodriguez97} and differs slightly from their definition.  We like to 
note that Eq.~(\ref{Kbin}) for $K(q)$ is not only valid for positive $q$ 
values, but also for negative $q$ values.  This feature allows us to 
determine numerically the full singularity spectrum $f(\alpha)$.  In the 
analysis we focused on long time scales, excluding the crossover regime, 
and used detrending methods.  We consider it as particularly interesting 
that only two parameters $a$ and $b$ or, equivalently, $H$ and $\Delta 
\alpha$, are sufficient to describe $\tau(q)$ and $K(q)$ for positive as 
well as negative $q$ values.  This strongly supports the idea of 'universal' 
multifractal behaviour of river runoffs as suggested (in different context) 
by Lovejoy and Schertzer. 

It is interesting to note that the generalized fluctuation functions we 
studied do not show any kind of multifractal phase transition at some 
critical value $q_D$ in the $q$-regime ($-10 \le q\le 10$) we analysed. 
Instead, our analysis shows a crossover at a specific time scale $s_\times$ 
(typically weeks) that weakly increases with decreasing moment $q$. In this 
paper, we concentrated on the large-time regime ($s\gg s_\times$), 
where we obtained coherent multifractal behaviour and did not see any 
indication for a multifractal phase transition. But this does not exclude 
the possibility that at small scales a breakdown of mulifractality at a 
critical $q$-value may occur, as has been emphasized by \citet{Tessier96} 
and \citet{Pandey98}.

\subsection{Extended Multiplicative Cascade Model} \label{model}

In the following, we like to motivate the 2-parameter formula
Eq.~(\ref{Hbin2}) and show how it can be obtained from the well known
multifractal cascade model \citep{Feder88,Barabasi91,Kantelhardt02}.  In 
the model, a record $\phi_k$ of length $N=2^{n_{\rm max}}$ is constructed 
recursively as follows:  
In generation $n=0$, the record elements are constant, i.e.  $\phi_k = 1$ for 
all $k=1, \ldots, N$.  In the first step of the cascade (generation $n=1$), the 
first half of the series is multiplied by a factor $a$ and the second half of 
the series is multiplied by a factor $b$.  This yields $\phi_k=a$ for $k=1, 
\ldots, N/2$ and $\phi_k=b$ for $k=N/2+1, \ldots, N$. The parameters $a$ and 
$b$ are between zero and one, $0 < a < b < 1$. Note that we do not restrict 
the model to $b = 1-a$ as is often done in the literature \citep{Feder88}.  
In the second step (generation $n=2$), we 
apply the process of step 1 to the two subseries, yielding $\phi_k=a^2$ for 
$k=1, \ldots, N/4$, $\phi_k=ab$ for $k=N/4+1, \ldots, N/2$, $\phi_k=ba=ab$ for 
$k=N/2+1, \ldots, 3N/4$, and $\phi_k=b^2$ for $k=3N/4+1, \ldots, N$.  In 
general, in  step $n+1$, each subseries of step $n$ is divided into two 
subseries of equal length, and the first half of the $\phi_k$ is multiplied 
by $a$ while the second half is multiplied by $b$.  For example, in generation 
$n=3$ the values in the eight subseries are $a^3, \; a^2b, \; a^2b, \; ab^2, 
\; a^2b, \; ab^2, \; ab^2, \;b^3$.  After $n_{\rm max}$ steps, the final 
generation has been reached, where all subseries have length 1 and no more 
splitting is possible.  We note that the final record can  be written as
$\phi_k = a^{n_{\rm max}-n(k-1)} b^{n(k-1)}$, where $n(k)$ is the number of 
digits 1 in the binary representation of the index $k$, e.~g. $n(13) = 3$, 
since 13 corresponds to binary 1101.

For this multiplicative cascade model, the formula for $\tau(q)$ has been 
derived earlier \citep{Feder88,Barabasi91,Kantelhardt02}.  The result is
$\tau(q) = [-\ln(a^q + b^q) +q \ln(a+b)]/ \ln 2$ or
\begin{equation} h(q) = {1 \over q} - {\ln(a^q + b^q) \over q\ln 2} + 
{\ln(a+b) \over \ln 2}. \label{Hbin1} \end{equation}
It is easy to see that $h(1)=1$ for all values of $a$ and $b$.  Thus, in 
this form the model is limited to cases where $h(1)$, which is the exponent 
Hurst defined originally in the $R/S$ method, is equal to one.  In order 
to generalize this multifractal cascade process such that any value of 
$h(1)$ is possible, we have subtracted the offset $\Delta h = \ln(a+b) / 
\ln(2)$ from $h(q)$.  The constant offset $\Delta h$ corresponds to additional 
long-term correlations incorporated in the multiplicative cascade model.  
For generating records without this offset, we rescale the power spectrum.  
First, we fast-Fourier transform (FFT) the simple multiplicative cascade data 
into the frequency domain.  Then, we multiply 
all Fourier coefficients by $f^{-\Delta h}$, where $f$ is the frequency.  
This way, the slope $\beta$ of the power spectra $E(f) \sim f^{-\beta}$ 
(the squares of the Fourier coefficients) is decreased from $\beta = 2 h(2) 
- 1 = [2 \ln(a+b) - \ln(a^2 + b^2)] /\ln 2$ into $\beta' = 2 [h(2) - \Delta h] 
- 1 = -\ln(a^2 + b^2) /\ln 2$, which is consistent with Eq.~(\ref{Hbin2}).  
Finally, backward FFT is employed to transform the signal back into the time 
domain.  A similar Fourier filtering technique has been used by 
\citet{Tessier96} when generating surrogate runoff data.

\subsection{Comparison with Model Data} \label{surrogate}

\begin{figure}
\noindent\centering\epsfig{file=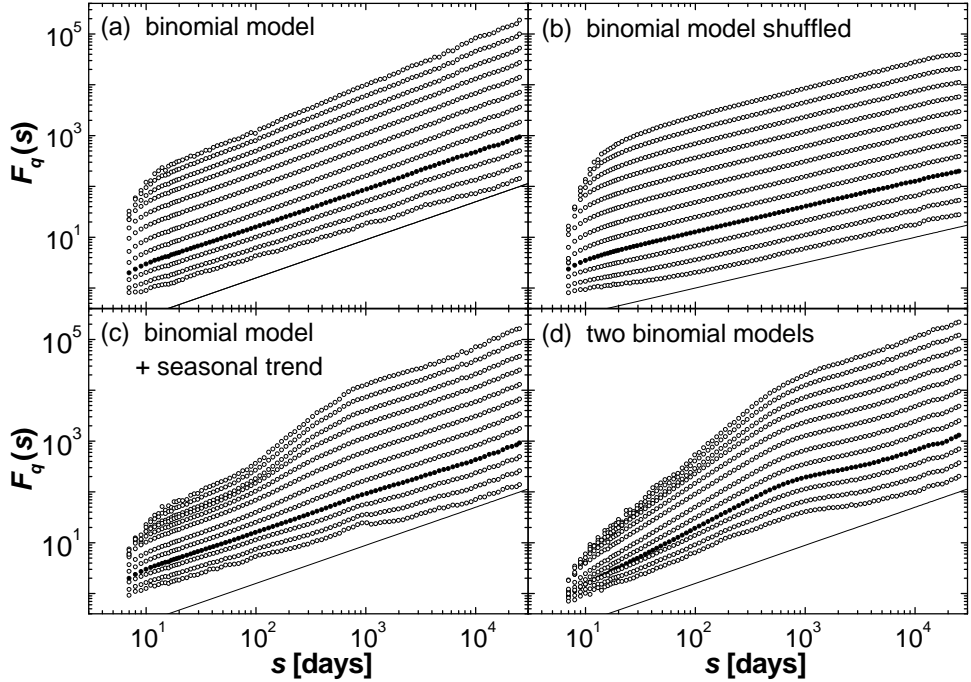,width=9cm,angle=-90}
\caption{The fluctuation functions $F_q(s)$ obtained from the multifractal 
DFA4 for surrogate series generated by the extended multiplicative cascade
model with parameters $a=0.50$ and $b=0.68$, that correspond to 
the values we obtained for the river Weser.  The fluctuation function 
$F_q(s)$ for (a) the original $\phi_i$ series and (b) the shuffled 
series are plotted versus scale $s$ for the same values of $q$ as in 
Figs. 4 and 5.  In (c) the $\phi_i$ have been multiplied by $0.1 + 
\sin^2(\pi i/365)$ before the analysis to simulate a seasonal trend.  In 
(d) modified values of the parameters $a$ and $b$ ($a=0.26$, $b=0.59$) 
have been used on scales $s \le 256$ to simulate the apparent stronger 
multifractality on smaller scales observed for most rivers.  For the figure, 
results from 10 surrogate series of length 140 years were averaged.}
\end{figure}

In order to see how well the extended multiplicative cascade model fits 
to the real data (for a given river), we generate the model data as follows:  
(i) we determine $a$ and $b$ for the given river (by best fit of 
Eq.~(\ref{Hbin2})), (ii) we generate the simple multiplicative cascade model 
with the obtained $a$ and $b$ values, and (iii) we implement the proper 
long-term correlations as described above.

Figure 8(a) shows the DFA analysis of the model data with parameters $a$ 
and $b$ determined for the river Weser.  By comparing with Fig.~4(a) we see 
that the extended model gives the correct scaling of the fluctuation 
functions $F_q(s)$ on time scales above the crossover.  By comparing 
Fig.~8(b) with Fig.~4(c) we see that the shuffled model series becomes 
uncorrelated without multifractality similar to the shuffled data.  Below the 
crossover, however, the model does not yield the observed $F_q(s)$ in the 
original data.  In the following we show that in order to obtain the proper 
behaviour below the crossover, either seasonal trends that cannot be 
completely eliminated from the data or a different type of multifractality 
below the crossover, represented by different values of $a$ and $b$, have 
to be introduced.  

To show the effect of seasonal trends, we have multiplied the elements 
$\phi_i$ of the extended cascade model by $0.1 + \sin^2(\pi i/365)$, this 
way generating a seasonal trend of period 365 in the variance.  
Figure 8(c) shows the DFA4 result for the generalized fluctuation functions, 
which now better resembles the real data than Fig. 8(a).  Finally, in Fig. 
8(d) we show the effect of a different multifractality below the crossover, 
where different parameters $a$ and $b$ characterize this regime.  The results 
also show better agreement with the real data. When comparing Figs.~8(a,c,d) 
with Figs.~4 and 5, it seems that the Danube, Amper and Wertach fit better to 
Fig.~8(d), i.e. suggesting different multifractality for short and large 
time scales, while the Weser, Susquehanna and Niger fit better to Fig.~8(c) 
where seasonal trends in the variance (and possibly in the skew)  
are responsible for the behaviour below the crossover.

\section{Conclusion}

In this study, we analyzed the scaling behaviour of daily runoff 
time series of 18 representative rivers in southern Germany and 23 
international rivers using both Detrended Fluctuation Analysis 
 and wavelet techniques.  In all cases we found that the 
fluctuations exhibit self-affine scaling behaviour and long-term 
persistence on time scales ranging from weeks to decades.  The 
fluctuation exponent $H$ varies from river to river in a wide range 
between 0.55 and 0.95, showing non-universal scaling behaviour.  

We also studied the multifractal properties of the runoff time series 
using a multifractal generalization of the DFA method that was 
crosschecked with the WTMM technique.  We found that the multifractal 
spectra of all 41 records can be  
described by a 'universal' function $\tau(q) = -\ln (a^q+b^q)/\ln 2$, 
which can be obtained from a generalization of the multiplicative cascade 
model and has solely two parameters $a$ and $b$ or, equivalently, 
the fluctuation exponent $H = {1 \over 2} - \ln (a^2 + b^2) /\ln 4$ 
and the width $\Delta \alpha = \ln {a \over b} / \ln 2$ of the 
singularity spectrum.  Since our function for $\tau(q)$ applies also 
for negative $q$ values, we could derive the singularity spectra 
$f(\alpha)$ from the fits.  We have calculated and listed the values 
of $H$, $a$, $b$, and $\Delta \alpha$ for all records considered.  
There are no significant differences between their distributions for 
rivers in South Germany and for international rivers. We also found that 
there is no significant dependence of these parameters on the size of the 
basin area, but there is a slight decrease of the multifractal width 
$\Delta \alpha$ with increasing basin area.  We suggest that the values 
of $H$ and $\Delta \alpha$ can be regarded as 'fingerprints' for each 
station or river, which can serve as an efficient non-trivial test bed 
for the state-of-the-art precipitation-runoff models.

Apart from the practical use of Eq.~(\ref{Hbin2}) with the parameters 
$a$ and $b$ that was derived by extending the multiplicative cascade model 
and that can serve as finger prints for the river flows, we presently are 
lacking a physical model for this behaviour.  It will be interesting to 
see, if physically based models, e.g. the random tree-structure model 
presented in \citet{Gupta96}, can be related to the multiplicative 
cascade model presented here. If so, this would give a physical explanation 
for how the multiplicative cascade model is able to simulate river flows.

We have also investigated the origin of the multifractal scaling 
behaviour by comparison with the corresponding shuffled data.  We 
found that the multifractality is removed by shuffling that destroys 
the time correlations in the series while the distribution of the runoff 
values is not altered.  After shuffling, we obtain $h(q) \approx 1/2$ 
for all values of $q$, indicating monofractal behaviour.  Hence, our 
results suggest that the multifractality is not due to the existence 
of a broad, asymmetric (singular) probability density distribution 
\citep{Anderson98}, but due to a specific dynamical 
arrangement of the values in the time series, i.e. a self-similar 
'clustering' of time patterns of values on different time scales.  
We believe that our results will be useful also to improve the 
understanding of extreme values (singularities) in the presence of 
multifractal long-term correlations and trends. 

Finally, for an optimal water management in a given basin, it is essential 
to know, whether an observed long-term fluctuation in discharge data is 
due to systematic variations (trends) or the results of long-term 
correlation.  Our approach is also a step forward in this directions.

{\it Acknowledgments:} We would like to thank Daniel Schertzer and Diego
Rybski for valuable discussions.  This work was supported by the BMBF, the 
DAAD, and the DFG.  We also would like to thank the Water Management Authorities 
of Bavaria and Baden-W\"urttemberg (Germany), and the Global Runoff Center 
(GRDC) in Koblenz (Germany) for providing the observational data.


\begin{thebibliography}{}

\bibitem[Anderson and Meerschaert(2002)]{Anderson98} Anderson, P.L., 
Meerschaert, M.M., 1998. Modelling river flows with heavy tails. 
{\it Water Resources Research}, {\it 34}(9), 2271-2280.

\bibitem[Arneodo et al.(2002)]{Arneodo02} Arneodo, A., Audit, B., 
Decoster, N., Muzy, J.-F., Vaillant, C., 2002.  Wavelet based 
multifractal formalism: Applications to DNA sequences, satellite images 
of the cloud structure, and stock market data.  In: \citet{Bunde02}, 
pp. 27-102.

\bibitem[Barabasi and Vicsek(1991)]{Barabasi91} Barabasi, A., 
Vicsek, T., 1991.  Multifractality of self-affine fractals.
{\it Phys. Rev. A}, {\it 44}, 2730-2733.

\bibitem[Bunde et al.(2000)]{Bunde00} Bunde, A., Havlin, S., Kantelhardt, 
J.W., Penzel, T., Peter, J.-H., Voigt, K., 2000.  Correlated and 
uncorrelated regions in heart-rate fluctuations during sleep. {\it Phys. 
Rev. Lett.}, {\it 85}(17), 3736-3739.

\bibitem[Bunde et al.(2002)]{Bunde02} Bunde, A., Kropp, J., 
Schellnhuber, H.-J. (eds.), 2002.  The science of disaster: Climate 
disruptions, market crashes, and heart attacks. Springer, Berlin.

\bibitem[Davis et al.(1996)]{Davis96} Davis, A., Marshak, A., Wiscombe, W.,
Cahalan, R., 1996.  Multifractal characterization of intermittency in 
nonstationary geophysical signals and fields.  In: Current topics in 
nonstationary analysis, edited by Trevino, G., Harding, J., Douglas, B., 
Andreas, E., World Scientific, Singapore, pp. 97-158.

\bibitem[Eichner et al.(2003)]{Eichner03} Eichner, J. F., Koscielny-Bunde, 
E., Bunde, A., Havlin, S., Schellnhuber, H.-J., 2003.  Power-law persistence 
and trends in the atmosphere: A detailed study of long temperature records.
{\it Phys. Rev. E}, {\it 68}, 046133.

\bibitem[Feder(1988)]{Feder88} Feder, J., 1988. Fractals.  Plenum 
Press, New York.

\bibitem[Frisch and Parisi(1985)]{Frisch85} Frisch, U., Parisi, G., 
1985.  Fully developed turbulence and intermittency, in: Turbulency and 
predictability in geophysical fluid dynamics.  Edited by Ghil, M., 
Benzi, R., Parisi, G., North Holland, New York, pp. 84-92. 

\bibitem[Gupta et al.(1994)]{Gupta94} Gupta, V.K., Mesa, O.J., 
Dawdy, D.R., 1994.  Multiscaling theory of flood peaks: Regional 
quantile analysis.  {\it Water Resources Research}, {\it 30}(12), 
3405-3421.

\bibitem[Gupta and Dawdy(1995)]{Gupta95} Gupta, V.K., Dawdy, D.R., 
1995.  Physical Interpretations of regional variations in the scaling 
exponents of flood quantiles.  In: Kalma, J.D., Scale issues in 
hydrological modelling, Wiley, Chichester, pp. 106-119.  

\bibitem[Gupta et al.(1996)]{Gupta96} Gupta V.K., Castro, S.L., 
Over, T.M., 1996. On scaling exponents of spatial peak flows from 
rainfall and river network geometry.  {\it J. Hydrol.}, {\it 187}(1-2), 
81-104.

\bibitem[Hurst(1951)]{Hurst51} Hurst, H.E., 1951.  Long-term storage 
capacity of reservoirs.  {\it Transactions of the American Society of 
Civil Engineering}, {\it 116}, 770-799.

\bibitem[Hurst et al.(1965)]{Hurst65} Hurst, H.E., Black, R.P., 
Simaika, Y.M., 1965. Long-term storage: An experimental study.
Constable \& Co. Ltd., London.

\bibitem[Kantelhardt et al.(2001)]{Kantelhardt01} Kantelhardt, J.W., 
Koscielny-Bunde, E., Rego, H.H.A., Havlin, S., Bunde, A., 2001.
Detecting long-range correlations with detrended fluctuation analysis.
{\it Physica A}, {\it 295}, 441-454.

\bibitem[Kantelhardt et al.(2002)]{Kantelhardt02} Kantelhardt, J. W., 
Zschiegner, S.A., Koscielny-Bunde, E., Havlin, S., Bunde, A., Stanley, 
H.E., 2002.  Multifractal detrended fluctuation analysis of 
nonstationary time series.  {\it Physica A}, {\it 316}, 87-114.

\bibitem[Kantelhardt et al.(2003)]{Kantelhardt03} Kantelhardt, J. W., 
Rybski, D., Zschiegner, S.A., Braun, P., Koscielny-Bunde, E., Livina, 
V., Havlin, S., Bunde, A., 2003.  Multifractality of river runoff 
and precipitation: Comparison of fluctuation analysis and wavelet 
methods.  {\it Physica A}, {\it 330}, 240-245.

\bibitem[Koscielny-Bunde et al.(1998)]{Koscielny98} Koscielny-Bunde, E., 
Bunde, A., Havlin, S., Roman, H.E., Goldreich, Y., Schellnhuber, H.-J.,
1998.  Indication of a universal persistence law governing atmospheric 
variability. {\it Phys. Rev. Lett.}, {\it 81}(3), 729-732.   

\bibitem[Lavallee et al.(1993)]{Lavallee93} Lavallee, D., Lovejoy, S., 
Schertzer, D., 1993.  Nonlinear variability and landscape topography: 
analysis and simulation.  In: {\it Fractals in Geography. PTR Prentic-Hall}, 
edited by DeCola, L., Lam, N., pp. 158-192, 1993.

\bibitem[Livina et al.(2003a)]{Livina03a} Livina, V. N., Ashkenazy, Y., 
Braun, P., Monetti, R., Bunde, A., Havlin, S., 2003a.  Nonlinear volatility 
of river flux fluctuations. {\it Phys. Rev. E, 67}, 042101.

\bibitem[Livina et al.(2003b)]{Livina03b} Livina, V., Ashkenazy, Y., 
Kizner, Z., Strygin, V., Bunde, A., Havlin, S., 2003b.  A stochastic model 
of river discharge fluctuations, {\it Physica A}, {\it 330}, 283-290.

\bibitem[Lovejoy and Schertzer(1991)]{Lovejoy91} Lovejoy, S., Schertzer, D.,
1991.  Nonlinear Variability in Geophysics: Scaling and Fractals.  Kluver 
Academic Publ., Dortrecht, Netherlands.

\bibitem[Mandelbrot and Wallis(1968)]{Mandelbrot68} Mandelbrot, B. B., 
Wallis, J. R., 1968.  Noah, Joseph, and operational hydrology. {\it Water 
Resources Research}, {\it 4}(5), 909.

\bibitem[Mandelbrot and Wallis(1969)]{Mandelbrot69} Mandelbrot, B. B., 
Wallis, J. R., 1969.  Some long-run properties of geophysical records.
{\it Water Resources Research}, {\it 5}(2), 321-340.

\bibitem[Matsoukas et al.(2000)]{Matsoukas00} Matsoukas C., Islam, S., 
Rodriguez-Iturbe, I., 2000. Detrended fluctuation analysis of rainfall 
and streamflow time series.  {\it J. Geophys. Res. Atmosph.}, 
{\it 105}(D23), 29165-29172.

\bibitem[Montanari et al.(2000)]{Montanari00} Montanari, A., Rosso, R., 
Taqqu, M. S., 2000.  A seasonal fractional ARIMA model applied to the Nile 
River monthly flows at Aswan. {\it Water Resources Research}, {\it 36}(5), 
1249-1259.

\bibitem[Pandey et al.(1998)]{Pandey98} Pandey, G., Lovejoy, S., 
Schertzer, D., 1998.  Multifractal analysis of daily river flows 
including extremes for basins of five to two million square kilometers, 
one day to 75 years.  {\it Journal of Hydrology}, {\it 208}, 62-81.

\bibitem[Peng et al.(1994)]{Peng94} Peng, C.-K., Buldyrev, S. V., 
Havlin, S., Simons, M., Stanley, H. E., Goldberger, A. L., (1994). 
Mosaic organization of DNA nucleotides, {\it Phys. Rev. E},{\it 49}(2), 
1685-1689.

\bibitem[Peters et al.(2002)]{Peters02} Peters, O., Hertlein, C., 
Christensen, K., 2002.  A Complexity View of Rainfall.  {\it Phys. Rev. 
Lett.}, {\it 88}, 018701. 

\bibitem[Rodriguez-Iturbe and Rinaldo(1997)]{Rodriguez97} 
Rodriguez-Iturbe, I., Rinaldo, A., 1997.  Fractal River Basins -- Change 
and Self-Organization.  Cambridge University Press, Cambridge.

\bibitem[Schertzer and Lovejoy(1987)] {Schertzer87} Schertzer, D., 
Lovejoy, S., 1987. Physical modelling and analysis of rain and clouds 
by anisotropic scaling multiplicative processes. {\it J. Geophys. Res. 
Atmosph.}, {\it 92}, 9693. 

\bibitem[Sivapalan et al.(2002)]{Sivapalan02} Sivapalan, M., Jothityangkoon, 
C., Menabde, M., 2002.  Linearity and nonlinearity of basin response as 
a function of scale: Discussion of alternative definitions.  {\it Water 
Resour. Res.}, {\it 38}, 1012.

\bibitem[Talkner and Weber(2000)]{Talkner00} Talkner, P., Weber, R. O., 2000. 
Power spectrum and detrended fluctuation analysis: Application to daily 
temperatures.  {\it Phys. Rev. E}, {\it 62}(1), 150-160.

\bibitem[Taqqu et al.(1995)]{Taqqu95} Taqqu, M. S., Teverovsky, V., 
Willinger, W., 1995.  Estimators for long-range dependence: An empirical 
study.  Fractals {\bf 3}, 785-798.  

\bibitem[Tessier et al.(1996)]{Tessier96} Tessier, Y., Lovejoy, S., Hubert, 
P., Schertzer, D., Pecknold, S., 1996.  Multifractal analysis and modelling 
of rainfall and river flows and scaling, causal transfer functions.  
{\it J. Geophys. Res. Atmosph.}, {\it 101}(D21): 26427-26440.

\bibitem[Turcotte and Greene(1993)]{Turcotte93} Turcotte, D. L., 
Greene, L., 1993.  A scale-invariant approach to flood-frequency analysis. 
{\it Stoch. Hydrol. Hydraul.}, {\it 7}, 33-40. 

\bibitem[Weber and Talkner(2001)]{Weber01} Weber, R. O., Talkner, P., 2001.
Spectra and correlations of climate data from days to decades. {\it 
J. Geophys. Res. Atmosph.}, {\it 106}(D17), 20131.

\end{thebibliography}
\end{document}